\documentclass[8.5pt,twoside,twocolumn]{article}
\usepackage[super,sort&compress,comma]{natbib} 
\usepackage{mhchem}
\usepackage{times,mathptmx}
\usepackage{sectsty}
\usepackage{balance} 

\usepackage{graphicx} 
\usepackage{lastpage}
\usepackage[format=plain,justification=raggedright,singlelinecheck=false,font=small,labelfont=bf,labelsep=space]{caption}

\usepackage{fancyhdr}
\pdfoutput=1

\begin{document}

\thispagestyle{plain}
\fancypagestyle{plain}{
\renewcommand{\headrulewidth}{1pt}}
\renewcommand{\thefootnote}{\fnsymbol{footnote}}
\renewcommand\footnoterule{\vspace*{1pt}%
\hrule width 3.4in height 0.4pt \vspace*{5pt}}
\setcounter{secnumdepth}{5}

\makeatletter
\def\subsubsection{\@startsection{subsubsection}{3}{10pt}{-1.25ex plus -1ex minus -.1ex}{0ex plus
0ex}{\normalsize\bf}}
\def\paragraph{\@startsection{paragraph}{4}{10pt}{-1.25ex plus -1ex minus -.1ex}{0ex plus
0ex}{\normalsize\textit}}

\renewcommand\@biblabel[1]{#1}
\renewcommand\@makefntext[1]%
{\noindent\makebox[0pt][r]{\@thefnmark\,}#1}
\makeatother
\renewcommand{\figurename}{\small{Fig.}~}
\sectionfont{\large}
\subsectionfont{\normalsize}

\renewcommand{\headrulewidth}{1pt}
\renewcommand{\footrulewidth}{1pt}
\setlength{\arrayrulewidth}{1pt}
\setlength{\columnsep}{6.5mm}
\setlength\bibsep{1pt}

\makeatletter
\DeclareRobustCommand\onlinecite{\@onlinecite}
\def\@onlinecite#1{\begingroup\let\@cite\NAT@citenum\citealp{#1}\endgroup}
\makeatother

\twocolumn[
  \begin{@twocolumnfalse}
\noindent\LARGE{\textbf{
Ultrahigh Energy Density Li-ion Batteries Based
on Cathodes of 1D Metals with -Li-N-B-N- Repeating Units 
in $\alpha$-Li$_{x}$BN$_{2}$ (1$\leq$x$\leq$3) 
}}
\vspace{0.6cm}

\noindent\large{\textbf{K\'aroly N\'emeth$^{\ast}$\textit{$^{a}$}}
}\vspace{0.5cm}

\vspace{0.6cm}

\noindent \normalsize{
Ultrahigh energy density batteries based on $\alpha$-Li$_{x}$BN$_{2}$ (1$\leq$x$\leq$3) positive electrode
materials are predicted using
density functional theory calculations. The utilization of the reversible LiBN$_{2}$ +
2 Li$^{+}$ + 2 e$^{-}$ $\rightleftharpoons$ Li$_{3}$BN$_{2}$ electrochemical cell reaction
leads to a voltage of 3.62 V (vs Li/Li$^{+}$), theoretical energy densities of 3251 Wh/kg and 5927 Wh/L, 
with capacities of 899 mAh/g  and 1638 mAh/cm$^{3}$, while the cell volume of $\alpha$-Li$_{3}$BN$_{2}$
changes only 2.8 \% per two-electron transfer. These values are far superior to the best existing or
theoretically designed intercalation or conversion-based positive electrode materials.  
For comparison, the theoretical energy density of a Li-O$_{2}$/peroxide battery 
is 3450 Wh/kg (including the weight of O$_{2}$), that of a Li-S battery is 2600 Wh/kg, 
that of Li$_{3}$Cr(BO$_{3}$)(PO$_{4}$) (one of the best designer 
intercalation materials) is 1700 Wh/kg, while 
already commercialized LiCoO$_{2}$ allows for 568 Wh/kg.
$\alpha$-Li$_{3}$BN$_{2}$ is also known as a good Li-ion conductor
with experimentally observed 3 mS/cm ionic conductivity and 78 kJ/mol ($\approx$ 0.8 eV) 
activation energy of conduction. The attractive features of $\alpha$-Li$_{x}$BN$_{2}$
(1$\leq$x$\leq$3) are based on a crystal lattice of 1D conjugated polymers with -Li-N-B-N- repeating units.
When some of the Li is deintercalated from $\alpha$-Li$_{3}$BN$_{2}$ the crystal becomes a metallic
electron conductor, based on the underlying 1D conjugated $\pi$ electron system. Thus
$\alpha$-Li$_{x}$BN$_{2}$ (1$\leq$x$\leq$3) represents a new type of 1D conjugated polymers with great
potential for energy storage and other applications.
}
\vspace{0.5cm}
 \end{@twocolumnfalse}
  ]

\section{Introduction} \label{introduction}
There is a quest for materials that would allow for storing large amounts of energy per unit weight and
volume and can be utilized as electroactive species in electrochemical energy storage devices.
These materials typically serve as part of the positive electrode of batteries while negative
electrodes may be composed of bulk metals, such as Li, Na, Mg, Al, etc, or as electrically and
ionically conductive composite materials containing these metals. 
The driving force of the discharge process in batteries
is the chemical potential difference of electrons and mobile cations in the positive and negative
electrodes manifesting as voltage between the current collectors. 
Main stream battery research focuses on improving Li-ion batteries that are based on positive
electrode materials capable of intercalating/deintercalating Li$^{+}$ ions during the charge/discharge
processes. The most promising Li$^{+}$ ion intercalating materials involve layered oxides LiMO$_{2}$,
spinels LiM$_{2}$O$_{4}$, polyanionic compounds such as 
olivine compounds LiMPO$_{4}$, silicate compounds Li$_{2}$MSiO$_{4}$, tavorites
LiMPO$_{4}$F and borates LiMBO$_{3}$, where M denotes a transition metal atom, usually Co, Mn, Fe,
Cr, V, Ni and Ti, sometimes M=Al may also occur \cite{BXu12,BCMelot13}. There are also conversion based
cell reactions when the mobile cation does not intercalate into a host crystal but forms separate
crystals with anions taken out from the host material. Examples of conversion-based batteries include
the reactions of Li with S \cite{MKSong13}, (CF)$_{x}$ \cite{PMeduri13} and FeF$_{3}$ \cite{LLi12}.
Representative theoretical gravimetric energy densities of these batteries are 
2600 Wh/kg for Li-S \cite{MKSong13}, 1950 Wh/kg for Li-FeF$_{3}$ \cite{LLi12}, 
1700 Wh/kg for Li$_{3}$Cr(BO$_{3}$)(PO$_{4}$) \cite{GHautier11} and 568 Wh/kg for LiCoO$_{2}$
\cite{MMThackeray12}. In practice, these values are far smaller due to additional materials that are
needed for practical implementations of the corresponding electrochemistries. For example, best
realized Li-S batteries allow for 300-500 Wh/kg \cite{MKSong13} which is still 
significantly better than that of commercially available batteries (130-200 Wh/kg, based on LiMO$_{2}$).
For intercalation-based cathode materials there is typically a factor of 3-4 difference between the
theoretical and practical energy density values.
Research is also conducted on metal-air type batteries that have extremely large theoretical energy
densities. A rechargeable Li-O$_{2}$/peroxide battery has a theoretical energy density of 11 kWh/kg 
when O$_{2}$ is taken from air, or 3450 Wh/kg when O$_{2}$ is carried within the battery
\cite{JChristensen12}. 
However, the realization of rechargeable metal-air batteries appears the most difficult endeavor among all
battery development directions \cite{JChristensen12,KNemeth14co2}.

\begin{figure}[tb!]
\resizebox*{3.4in}{!}{\includegraphics{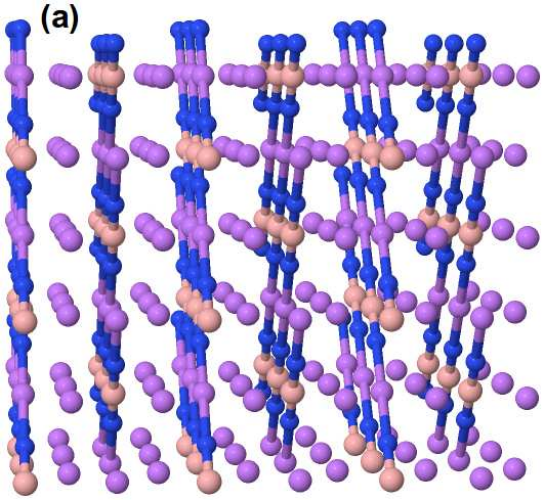}}
\resizebox*{3.4in}{!}{\includegraphics{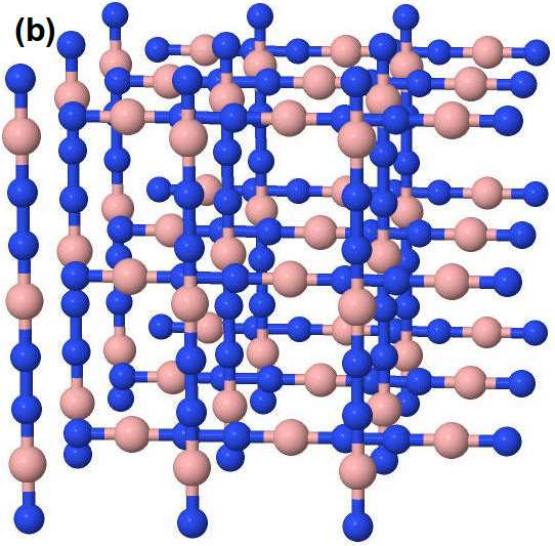}}
\caption{
The structure of $\alpha$-Li$_{3}$BN$_{2}$ (panel a) and that of the predicted derivative
$\alpha$-BN$_{2}$ compound (panel b), in 3x3x3 supercells. 
Color code: N - blue, Li - violet, B - magenta. Linear polymers with -Li-N-B-N- repeating units
are present in $\alpha$-Li$_{3}$BN$_{2}$, where they form layers. 
Between the layers, additional Li ions take place tetrahedrally coordinating to nearest N atoms. 
This latter Li ions are predicted to be deintercalatable without the collapse of the rest of the
crystal. When all Li-s are deintercalated, structures like $\alpha$-BN$_{2}$ may form.
}
\label{structures}
\end{figure}

As the ideal electrode material is both a good electron and ion conductor,
electrically conducting quasi 1D polymers have also been subject of battery research
\cite{novak1997electrochemically}. 
These polymers are based on conjugated $\pi$-electron systems, such as polyacetylene, 
polyaniline, polypyrrole, polythiophene, etc.
Perhaps the best performing polymer of this kind is polyaniline.
Polyaniline cathodes and lithium anodes can achieve a 
theoretical energy density of 340 Wh/kg (in reference to the weight of the electroactive materials) at
a voltage of 3.65 V \cite{novak1997electrochemically} using LiClO$_{4}$ electrolyte where the
ClO$_{4}^{-}$ ions play the role of dopant of the conductive polymer in the charged state of 
a supercapacitor type device whereby the polymer carries positive charges.  
Poly(sulfur nitride) (also called polythiazyl), the first known example of a polymeric conductor that is
also superconductor below 0.3 K, has also been tested
in battery applications, achieving 90 Wh/kg theoretical energy density \cite{archer2004inorganic}. 
However, due to its explosivity when heated to 240 $^{\rm o}$C, or due to mechanical impact or 
electrical ark \cite{downs1980electric} its application is rendered unsafe in batteries.   
Conjugated 1D polymers occur in many other materials, especially in coordination polymers, such as CuCN, AgCN,
AuCN or ternary acetylides \cite{ruschewitz2006ternary,terdik2012anomalous}. 

$\alpha$-Li$_{3}$BN$_{2}$ (space group P4$_{2}$/mnm \cite{cenzual1991inorganic}), 
a derivative of hexagonal boron nitride (h-BN) that forms when reacting h-BN with molten Li$_{3}$N
\cite{yamane1987high,devries1969system,goubeau1961ternare}, 
provides an interesting example of 1D conjugated polymers with -Li-N-B-N- repeating units.
The -N-B-N- part of the -Li-N-B-N- repeating unit is the dinitridoborate anion, BN$_{2}^{3-}$, thus
each repeating unit -Li-N-B-N- carries two negative charge. The two negative charge of the
-Li-N-B-N- repeating unit is counterbalanced by two Li$^{+}$ ions per formula unit,
located between sheets of the polymeric strands as shown in Fig. \ref{structures} a.
This layered structure strongly resembles to that seen for example in layered
oxide materials used in Li ion batteries.

In each polymer-containing layer, the polymer strands are running 
parallel with half the length of the repeating
unit shifted relative to the neighboring strand so that each B atom will have a Li neighbor 
in the neighboring strand. Nearest neighbor polymer-containing layers are 
rotated by 90 degrees relative to each other 
and are placed such that in the direction perpendicular to the layers 
each B atom will have a Li neighbor again. As a result, in
$\alpha$-Li$_{3}$BN$_{2}$ each B atom is octahedrally coordinated to 
neighboring Li atoms of polymeric strands,
and vice versa for the Li-atoms in the polymers. The distance between Li atoms and their nearest
N neighbors is 1.95 {\AA} in the polymers, which counts as a strong 
coordinative Li-N bond. The Li atoms between
the polymer-containing layers are coordinated tetrahedrally to four 
nearest N atoms of polymers at a distance of 2.12 {\AA}.
$\alpha$-Li$_{3}$BN$_{2}$ contains one two-coordinated Li atom (in the polymer) 
and two four-coordinated Li atoms
(between the polymeric layers) per formula unit. These Li atoms will be referred to as Li(2N) and Li(4N),
respectively, in the following.

There are two other known phases of Li$_{3}$BN$_{2}$ besides the $\alpha$ one.
A closely related other phase is the $\beta$ one (space group I4$_{1}$/amd)
\cite{yamane1987high,pinkerton2006tetragonal}. In $\beta$-Li$_{3}$BN$_{2}$ the polymer strands are running
parallel in the layers without any relative shift, 
so B atoms have B neighbors in neighboring strands (and vice
versa for Li). In the neighboring layers, B atoms have one B and one Li neighbor.
This results in an asymmetric force-field and bends the polymer strands 
by 7-8 degrees at each B atom and by 16 degrees
at each Li(2N) atom. The third known phase of Li$_{3}$BN$_{2}$ 
is the monoclinic one \cite{yamane1986structure} (space group P2$_{1}$/c), this phase does not
contain linear chains of -Li-N-B-N- repeating units, all Li-s are of Li(4N) type.

The Li-ion conductivities of the $\alpha$ and $\beta$ Li$_{3}$BN$_{2}$ and monoclinic phases have been
measured at T = 400 K temperature, nearly 30 years ago, 
and have been found to be 3, 6 and 6 mS/cm with activation energies of
78, 64 and 64 kJ/mol, respectively \cite{yamane1987high,yamane1987preparation,yamane1986structure}. 
These Li ion conductivities count as
good and are comparable to that of other Li-ion battery intercalation cathode materials. 
For example, LiCoO$_{2}$ has Li-ion conduction and intercalation/deintercalation 
activation energies of 37-69 kJ/mol \cite{qiu2012electrochemical}. 
Despite the analogies of the structure of
Li$_{3}$BN$_{2}$ with known Li-ion battery cathode materials and to its good ionic conductivity,
to the best of our knowledge,
Li$_{3}$BN$_{2}$ has not been investigated yet as a positive electrode 
electroactive material, it has only been considered as
Li-ion conductor \cite{waechter2013coating} or as component of conversion 
based anode \cite{mason2011first} materials.
Li$_{3}$BN$_{2}$ is an attractive candidate for intercalation-based 
positive electrode electroactive material, 
as it is built only of light elements of the second row of the periodic table, therefore it is much
lighter per formula unit than the typical intercalation cathode materials that contain heavy
transition metals. This low weight per formula unit may result in high gravimetric energy density,
when the corresponding cell reaction is energetic enough.
Therefore, the present study focuses on theoretical calculations regarding the potential application of
Li$_{3}$BN$_{2}$ phases as positive electrode materials in batteries. While 
$\alpha$-Li$_{3}$BN$_{2}$ has already been proposed 
for use as a positive electrode material by the present author 
in a recent publication \cite{KNemeth2014ijqc} and some of its most 
important characteristics has been briefly discussed there, 
the present work provides a comprehensive theoretical analysis of the system.

\section{Methodology}
In the $\beta$-Li$_{3}$BN$_{2}$ phase the -Li-N-B-N- polymers are bent and the collapse of
the polymers can be expected when sufficient Li ions are deintercalated. The monoclinic phase has
no polymers and it appears difficult to define a stable structure after the deintercalation of
Li-ions.
Therefore, the following computations will focus on the electrochemical properties of
$\alpha$-Li$_{3}$BN$_{2}$, where stable crystal structure of -Li-N-B-N- polymers 
can be expected even after the deintercalation of the Li-ions from the Li(4N) positions.
As there are two Li(4N) atoms per formula unit in $\alpha$-Li$_{3}$BN$_{2}$, 
the proposed cell reaction is the following:
\begin{equation}\label{cellreaction}
LiBN_{2} + 2 Li^{+} + 2 e^{-} \rightleftharpoons Li_{3}BN_{2} .
\end{equation}

The intercalation electrode potential of the cell reaction, relative to the anode potential, 
can be calculated from the Gibbs free energy of the reaction, $\Delta G$, 
expressed in eV and divided by the number of electrons transferred. If the
entropy contribution is small (volume change is negligible), $\Delta G$ can be approximated 
as the change of electronic energy, $\Delta E$ during the reaction
\cite{GHautier11,FZhou04,AJain11}. In the case of the above reaction, $\Delta E$ can be expressed as
\begin{equation}\label{DeltaE}
\Delta E = E(Li_{3}BN_{2}) - E(LiBN_{2}) - 2 E(Li) ,
\end{equation}
where E(Li$_{3}$BN$_{2}$), E(LiBN$_{2}$) and E(Li) refer to the electronic energy of the 
corresponding crystals per formula unit,
for $\alpha$-Li$_{3}$BN$_{2}$, for the derivative $\alpha$-LiBN$_{2}$ obtained after deintercalating
all Li(4N)-s and for bulk metallic Lithium, respectively.
The open circuit voltage of the above cell reaction will be U = -$\Delta E$ / 2 , as two electrons
have been transferred. This U voltage is identical with the intercalation electrode potential of
LiBN$_{2}$/LiBN$_{2}^{2-}$ relative to the Li/Li$^{+}$ electrode.

The electronic energy change during the cell reaction is often calculated using the DFT(GGA)+U method
\cite{anisimov1991band},
in order to account for problems of DFT in predicting properties of transition metal compounds
\cite{GHautier11,FZhou04,AJain11}. Since there is no transition metal in the present cell reaction,
a simpler approach may be followed here, using the PBEsol \cite{PBE,PBEsol} exchange-correlation
functional to calculate electronic energies and optimum crystal structures.
The Quantum-Espresso \cite{giannozzi2009quantum} code is used with 
ultrasoft pseudopotentials (as provided with the code) 
in a plane-wave basis with 50 rydberg wavefunction cutoff. A 10x10x10 k-space grid is used to
discretize the electronic bands (unless otherwise noted). 
Optimum crystal structures have less than 1.d-4 rydberg/bohr
residual forces with residual pressure of less than 1 kbar on the unit cells.
Electronic energies refer to optimum structures unless otherwise noted.
The methodology has been validated on experimental data of 
lattice parameters and enthalpies of formation of $\alpha$-Li$_{3}$BN$_{2}$, Li$_{3}$N and h-BN. 
Enthalpies of formations were estimated as the change of electronic energy during the formation
of the compounds from the corresponding elements at T = 0 K, i.e. from crystalline Li and B and N$_{2}$ gas.

Experimental lattice parameters of $\alpha$-Li$_{3}$BN$_{2}$ are a = b = 4.6435 {\AA} and c =
5.2592 {\AA} \cite{cenzual1991inorganic}, while calculated ones 
are 4.6019 and 5.1317 {\AA}, respectively, an agreement within
1.0 and 2.5 \%, respectively. For Li$_{3}N$ (with a 6x6x6 k-space), 
experimental lattice parameters are a = b = 3.6373 {\AA} and 
c = 3.8703 {\AA} \cite{huq2007structural}, while calculated ones 
are a = b = 3.5549 {\AA} and c = 3.8088 {\AA}, an agreement within
2.3 and 1.6 \%, respectively. For h-BN (with a 6x6x6 k-space), 
a = b = 2.5039 {\AA} and c = 6.6612 {\AA} \cite{pease1952x}, while calculated ones
are a = b = 2.5032 {\AA} and c = 7.0191 {\AA}, respectively, an agreement within 0.03 and 5.3 \%,
respectively;
note that the latter discrepancy is due to the known inaccuracy of the PBEsol functional to describe
intermolecular interaction.

The experimental standard enthalpy of formation of $\alpha$-Li$_{3}$BN$_{2}$ is -534.5 ( $\pm$ 16.7) kJ/mol 
\cite{mchale1999energetics}, the calculated one is -512.0 kJ/mol (4.2 \% difference). For Li$_{3}$N,
the corresponding values are
-164.5 kJ/mol \cite{MWJrChase98} and -161.5 kJ/mol, respectively (1.9 \% difference). For h-BN, the
experimental value is -250.9 kJ/mol \cite{wise1966fluorine}, the calculated one is -260.9 kJ/mol (3.9 \% 
difference).

\begin{figure}[tb!]
\resizebox*{3.4in}{!}{\includegraphics{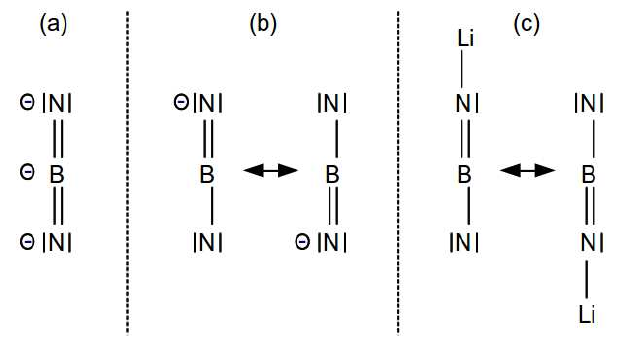}}
\caption{
Resonance structures of the dinitridoborate, BN$_{2}^{3-}$ anion (panel a) and its oxidized version,
BN$_{2}^{-}$ (panel b), as well as that of the -Li-N-B-N- repeating unit in LiBN$_{2}$.
}
\label{resonance}
\end{figure}

\section{Results and discussion}
The cell reaction in Eq. \ref{cellreaction} assumes the existence of the BN$_{2}^{-}$ anion, the 
oxidized form of BN$_{2}^{3-}$. While BN$_{2}^{3-}$ has been known for decades, no compounds with
its oxidized form, BN$_{2}^{-}$ are known, to the best of the present author's knowledge.

Resonance structures of the BN$_{2}^{3-}$ anion and related anions are discussed in Fig. 1 of
Ref. \onlinecite{Blaschkowski03}. Analogously, resonance structures of its oxidized form,
BN$_{2}^{-}$, have been depicted in the present work in Fig. \ref{resonance}. It appears that on
the basis of the resonance structures the existence of the oxidized form, BN$_{2}^{-}$ is possible,
especially when it is stabilized in linear chains with repeating BN$_{2}^{-}$ and Li$^{+}$ ions,
being part of a linear conjugated $\pi$-electron system.
The oxidation of BN$_{2}^{3-}$ can also be viewed as if BN$_{2}^{3-}$ would be a way of storing a nitride
ion, N$^{3-}$ absorbed to a neutral diatomic BN molecule and allowing for the electrochemistry of the 
N$^{3-}$ ion within this complex. The N$^{3-}$ ion represents an oxidation state of N as in ammonia
(NH$_{3}$) or in lithium nitride (Li$_{3}$N), while the oxidized form N$^{2-}$ is familiar from hydrazine 
(N$_{2}$H$_{4}$) and the further oxidized form N$^{-}$ is known from lithium-diazenide (Li$_{2}$N$_{2}$) 
\cite{schneider2012high}. The Li$_{3}$BN$_{2}$ molecule may be viewed as a combination of half of a
Li$_{2}$N$_{2}$ and half of a hydrazine-like Li$_{4}$N$_{2}$ through a B atom, whereby the B binds with a
double bond to the diazenide-type N and with a single bond to the hydrazine-type N.
Albeit oxidized forms of the BN$_{2}^{3-}$ ion, such as BN$_{2}^{2-}$ and BN$_{2}^{-}$ have not been
reported in the literature yet, there is recent indirect evidence to the existence of BN$_{2}^{2-}$ in
Na$_{2}$BN$_{2}$ obtained during thermogravimetric analysis of the thermolysis of Na$_{2}$KBN$_{2}$, 
after the evaporation of all K in a well separable step \cite{koz2014na3}.

The calculated intercalation electrode potential of the cell reaction in 
Eq. \ref{cellreaction} is 
U(LiBN$_{2}$/Li$_{3}$BN$_{2}$) = 3.62 V, relative to a Li/Li$^{+}$ electrode.
The energy of the cell reaction is $\Delta E$(LiBN$_{2}$/Li$_{3}$BN$_{2}$) = 7.24 eV per two
electron transfer. The mass-density of $\alpha$-Li$_{3}$BN$_{2}$ is $\rho$(Li$_{3}$BN$_{2}$) = 1.823 g/cm$^{3}$.
The gravimetric energy density is $\rho_{EG}$(Li$_{3}$BN$_{2}$) = 3251 Wh/kg,
the volumetric energy density is $\rho_{EV}$(Li$_{3}$BN$_{2}$) = 5927 Wh/L. The gravimetric
capacity, i.e. the concentration of de/re-intercalatable Li, is $\rho_{CG}$(Li$_{3}$BN$_{2}$) = 899
mAh/g, the volumetric one is  $\rho_{CV}$(Li$_{3}$BN$_{2}$) = 1638 mAh/cm$^{3}$ .
The cell volume of $\alpha$-Li$_{3}$BN$_{2}$ changes only by 2.8 \% per two-electron transfer (shrinks
during deintercalation).
Calculated lattice parameters of $\alpha$-Li$_{3}$BN$_{2}$ are 
a = b = 4.6019 and c = 5.1317 {\AA}, while for the deintercalated $\alpha$-LiBN$_{2}$ they are
a  = b = 4.5460 and c = 4.9613 {\AA}, note that each unit cell contains two formula units of material.
The calculated characteristic nearest atomic distances 
barely change during the deintercalation: the symmetric
Li(2N)-N, Li(4N)-N and B-N bond lengths are 1.912, 2.092 and 1.343 {\AA} in
$\alpha$-Li$_{3}$BN$_{2}$, they are 1.932, 2.062 and 1.337 {\AA} in $\alpha$-Li$_{2}$BN$_{2}$ and
1.888, none and 1.327 {\AA} in $\alpha$-LiBN$_{2}$, respectively.

The energy density and capacity values of $\alpha$-Li$_{3}$BN$_{2}$ 
are far superior to those of other known or designer
cathode materials, listed in Section \ref{introduction}. 
For a brief comparison, representative theoretical gravimetric energy densities of these other cell
reactions are 
2600 Wh/kg for Li-S (conversion-based) \cite{MKSong13}, 1950 Wh/kg for Li-FeF$_{3}$ (conversion-based) 
\cite{LLi12}, 
1700 Wh/kg for Li$_{3}$Cr(BO$_{3}$)(PO$_{4}$) (intercalation-based) \cite{GHautier11} 
and 568 Wh/kg for LiCoO$_{2}$ (intercalation-based) \cite{MMThackeray12}.
In fact, the theoretical energy density of $\alpha$-Li$_{3}$BN$_{2}$, 3251 Wh/kg, 
is very close (within 6 \%) to that of
a Li-O$_{2}$/peroxide battery, 3450 Wh/kg, when O$_{2}$ is carried within the battery
\cite{JChristensen12}.

\begin{figure}[tb!]
\resizebox*{3.3in}{!}{\includegraphics{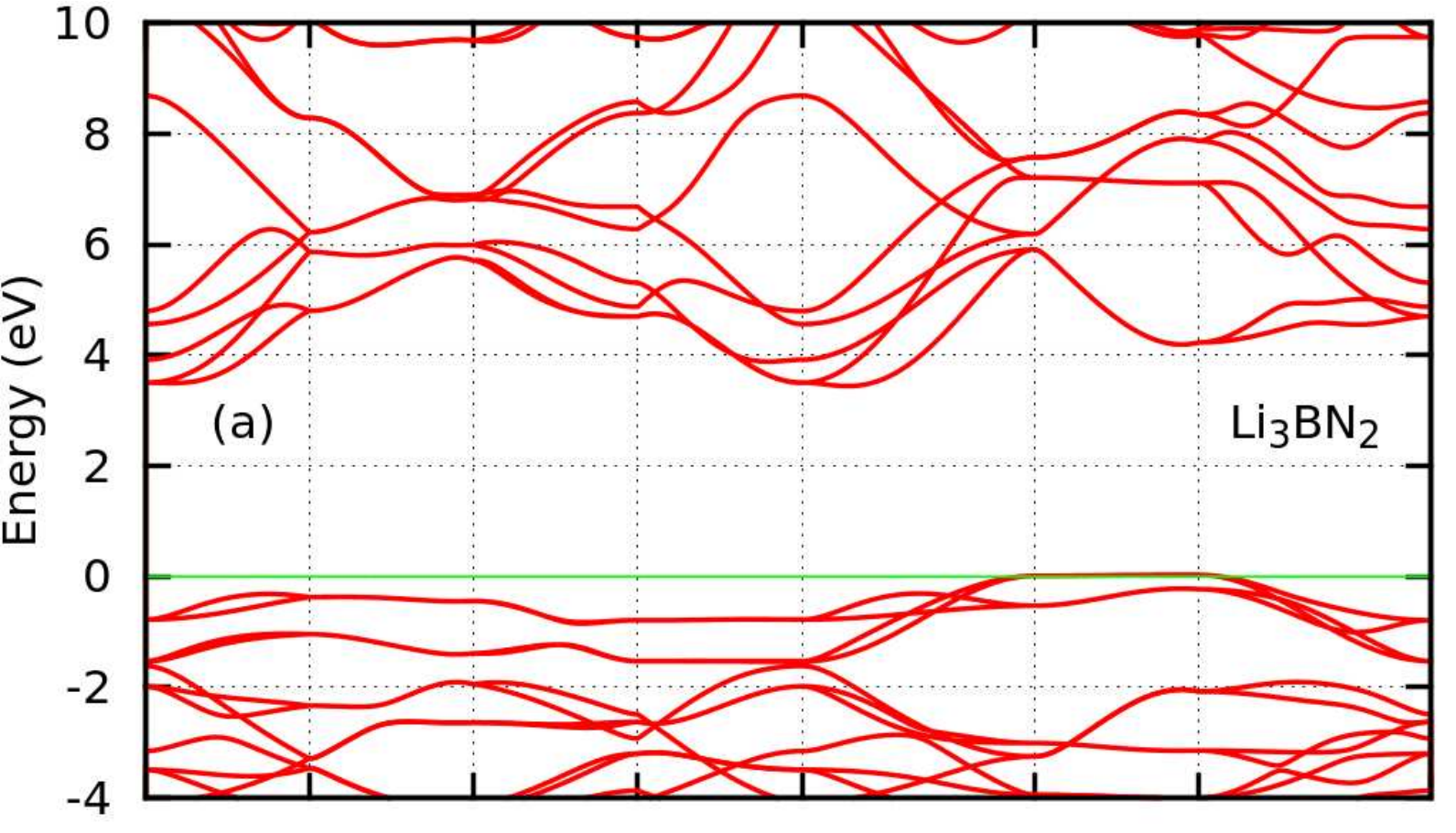}}
\resizebox*{3.3in}{!}{\includegraphics{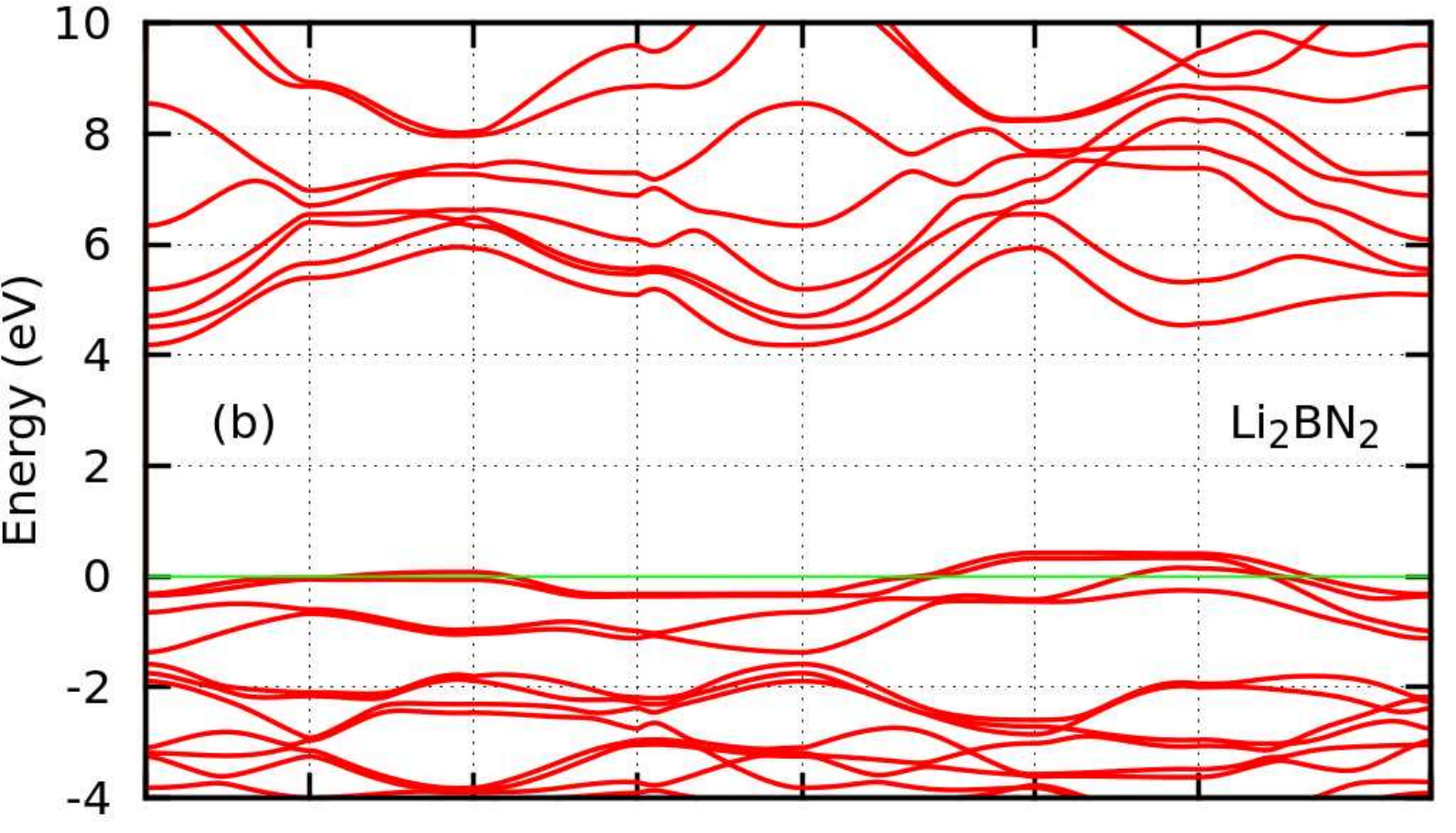}}
\resizebox*{3.3in}{!}{\includegraphics{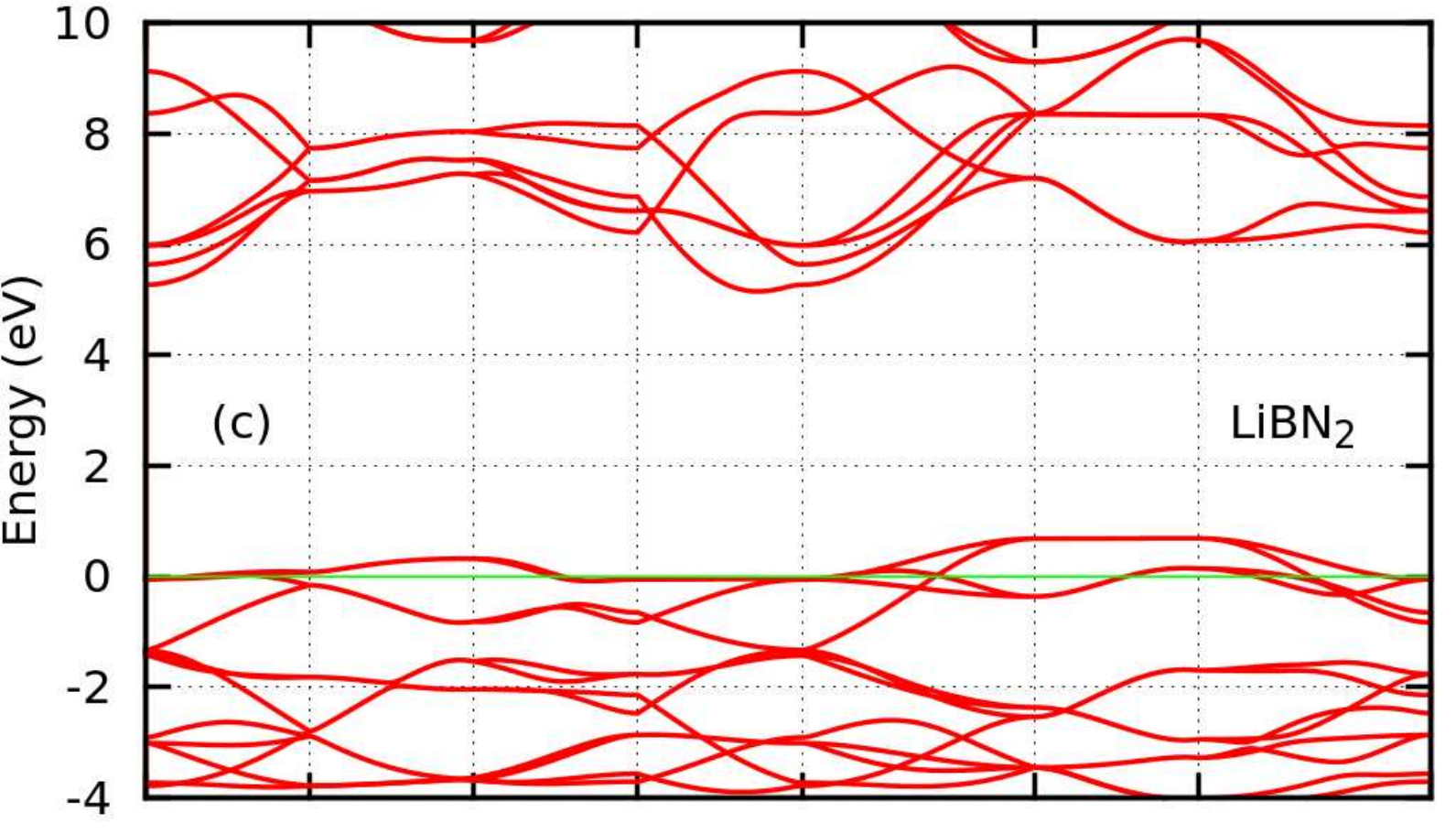}}
\resizebox*{3.3in}{!}{\includegraphics{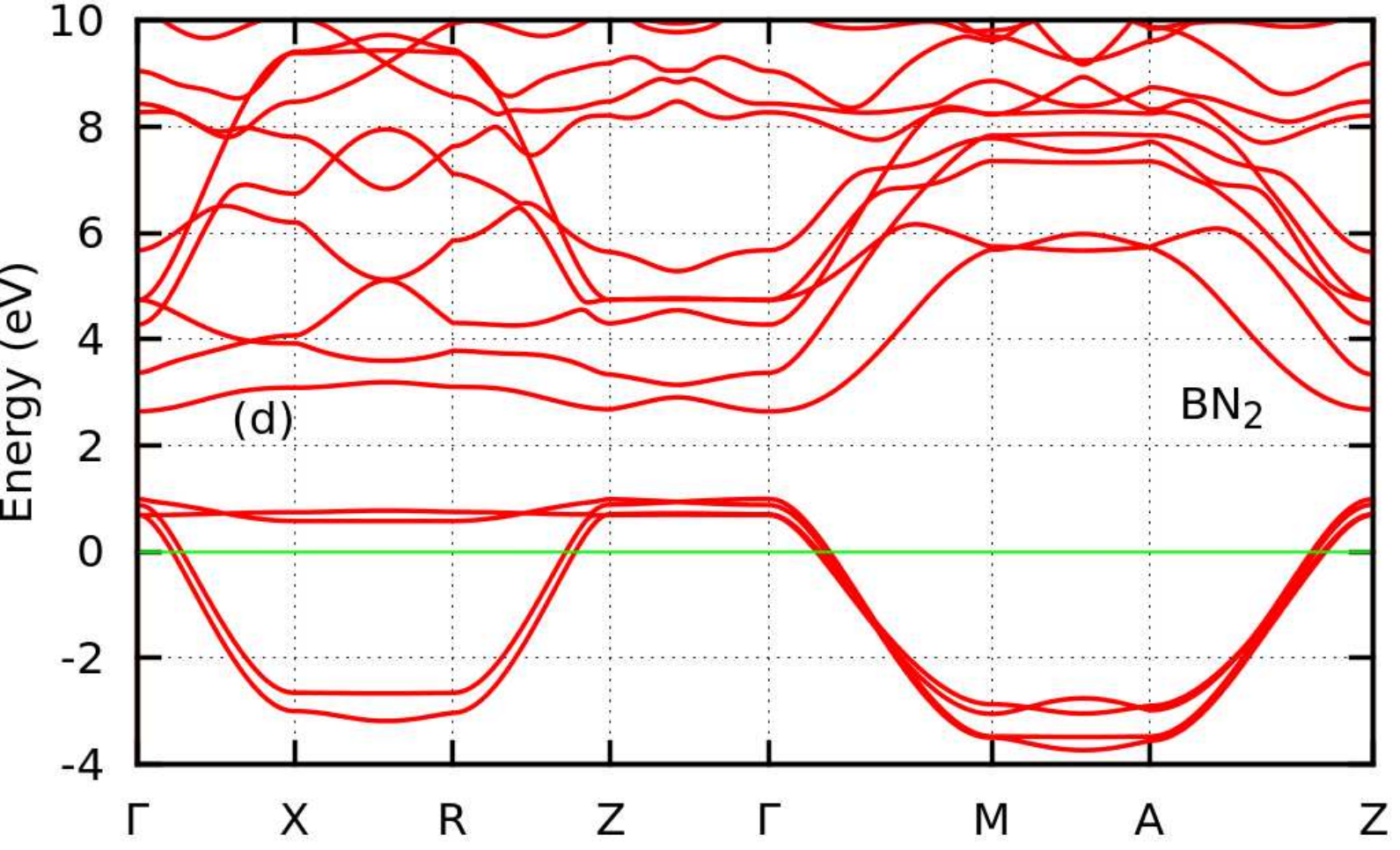}}
\caption{Bandstructures of compounds obtained by gradual extraction
of Li from $\alpha$-Li$_{3}$BN$_{2}$. 
The origin of the energy scale is set to the Fermi energy for each compound.
}
\label{bands}
\end{figure}

Fig. \ref{bands} shows the electronic bands in $\alpha$-Li$_{3}$BN$_{2}$ and in its gradually delithiated
versions, Li$_{2}$BN$_{2}$, LiBN$_{2}$ and BN$_{2}$. Note that the Li in the -Li-N-B-N- chains is
extracted only in the last step, in order to preserve the polymeric skeleton of the crystal. Many
possible structures may form after the complete extraction of the Li. 
From the many variants, the depicted structure of BN$_{2}$ in Fig. \ref{structures} (panel b) appears to be 
the most analogous one to $\alpha$-Li$_{3}$BN$_{2}$ in the sense that linear polymers with -N-B-N- repeating
units are preserved and the relative orientation of these polymers is similar to that in
$\alpha$-Li$_{3}$BN$_{2}$. As indicated by the bandstructures, the delithiation results in metallic
systems, as holes are created in the valence band of $\alpha$-Li$_{3}$BN$_{2}$ and the Fermi level
decreases below the top of the valence band of $\alpha$-Li$_{3}$BN$_{2}$ while no band-gap opens.
This phenomenon is similar to that observed in the LiCoO$_{2}$ cathode material which also becomes
metallic upon delithiation \cite{qiu2012electrochemical}. Only the 
$\alpha$-Li$_{x}$BN$_{2}$ (1$\leq$x$\leq$3) systems are suitable
for intercalation-based electrodes, as BN$_{2}$ has a significantly different cell volume.

The proposed $\alpha$-Li$_{x}$BN$_{2}$ (1$\leq$x$\leq$3) and BN$_{2}$ have merits also as new conductive polymers
based exclusively on Li, B and N or on B and N, opening up new paths in the field of synthetic metals
as well. 

Concerning the stability of the oxidized forms of BN$_{2}^{3-}$, i.e. that of BN$_{2}^{2-}$ and
BN$_{2}^{-}$, there is no experimental evidence yet, except the indirect one mentioned above for the
existence of BN$_{2}^{2-}$ in Na$_{2}$BN$_{2}$ \cite{koz2014na3}. The thermal decomposition of
Na$_{3}$BN$_{2}$ and Na$_{2}$KBN$_{2}$ as well as that of Na$_{2}$BN$_{2}$ leads to elemental alkali
metals, h-BN and N$_{2}$ at temperatures above 710 and 800 K \cite{koz2014na3}. At lower temperature
during electrochemical cycling such decomposition is not expected, 
as the binding of B to neighboring N is strong, also indicated by nearly identical bond lengths along
the -Li-N-B-N- repeating units in all $\alpha$-Li$_{x}$BN$_{2}$ (1$\leq$x$\leq$3). Thus the decomposition of
the oxidized forms of BN$_{2}^{3-}$ can be expected only at high temperature when the kinetic energy
of the vibration of the B-N bond is high enough to break the bond.

\begin{table}[tb!]
\caption{
Binding energies, $\varepsilon$, of Li(2N) and Li(4N) atoms in 
$\alpha$-Li$_{n}$BN$_{2}$ (n $\in$ [0,3]) 
and net charges Q on the various types of atoms.
Some charge is lost during the projection
of electron density from plane-wave basis into atomic orbital one (L\"owdin charges).
Binding energies for all systems have been calculated at the relaxed structure
obtained with the Li(2N) positions filled, except for $\varepsilon$(Li(2N)) at n=2
where the optimum structure of n=3 has been used. $\varepsilon$ values are 
relative to binding energy of Li in crystalline Li.
}
\label{bindingE}
\begin{tabular}{ccccccc}
\hline
    n    & \multicolumn{2}{c}{$\varepsilon$/eV} & \multicolumn{4}{c}{Q/e}  \\
         & Li(2N) & Li(4N) & B & N & Li(2N) & Li(4N) \\
\hline
    3    & -3.326  & -3.564  &  0.00  & -1.04  & +0.78 & +0.76 \\
    2    & -3.817  & -3.664  & +0.07  & -0.75  & +0.79 & +0.78 \\
    1    & -4.431  & -4.278  & +0.17  & -0.42  & +0.79 &   -   \\
    0    &    -    &    -    & +0.25  & -0.05  &   -   &   -   \\
\hline
\end{tabular}
\end{table}

\begin{figure}[bt!]
\resizebox*{3.4in}{!}{\includegraphics{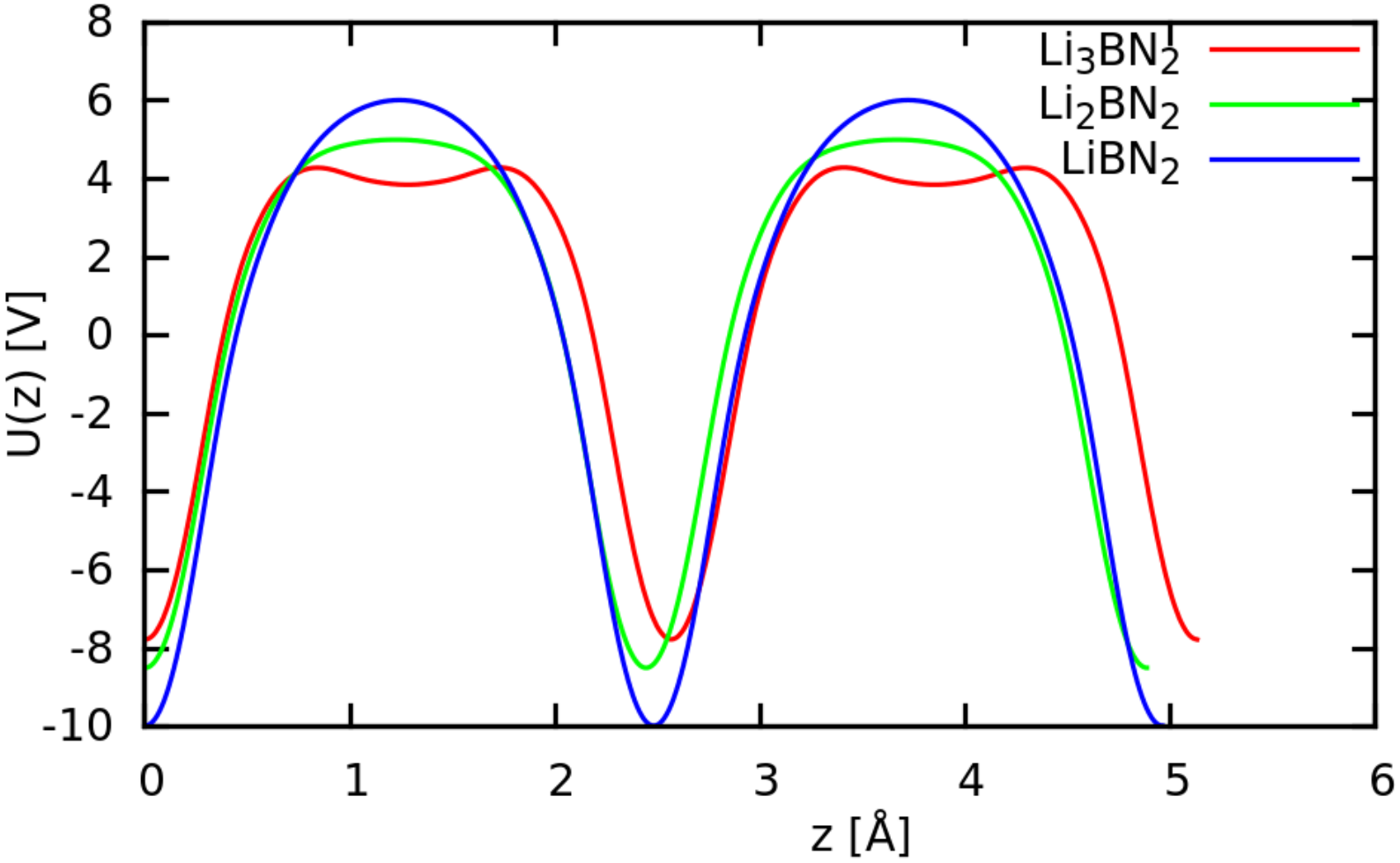}}
\caption{
The plane-averaged electrostatic potential, U(z) along the z axis (parallel with main crystallographic
axis, c) for
$\alpha$-Li$_{n}$BN$_{2}$, with n = 1, 2 or 3.
The minima are located in the planes filled with -Li-N-B-N- polymers.
The electrostatic potential attracts Li-ions to fill available Li-vacancies in
the -Li-N-B-N- chains, thereby self repairing -Li-N-B-N- chains during
the Li$_{3}$BN$_{2}$ $\rightarrow$ LiBN$_{2}$ + 2 Li deintercalation process,
i.e. on charging the battery.
The layers along the z axis of $\alpha$-Li$_{x}$BN$_{2}$ (1$\leq$x$\leq$3) act as
capacitor plates with alternating majority charges.
}
\label{potential}
\end{figure}

The binding energies of Li(2N) and Li(4N) atoms are listed in Table \ref{bindingE}, they also represent
the intercalation potentials of Li-s in the specific positions in the given optimum crystal structures.
Except for the completely lithiated crystal, in all other cases the Li(2N)-s are stronger bound than
the Li(4N)-s, i.e. the energetically favorable position of the Li ions is in the -Li-N-B-N- chains.
This result has been validated also in a 3x3x3 supercell of LiBN$_{2}$ (54 formula units, 216 atoms). At the
optimum geometry (obtained with all Li-s in Li(2N) positions) a single Li ion has been moved into a
tetrahedral Li(4N) position. As a result, the electronic energy of the system increased by 
${\Delta}E$(2N/4N) = 0.83 eV.
The Li(2N) filling appears more energetically favored over the Li(4N) one because the planes filled by
the -Li-N-B-N- polymers are negatively charged and the electrostatic potential has a minimum for Li$^{+}$
ions in these planes, as indicated in Fig. \ref{potential}.

The experimental value of the activation energy of Li-ion conduction in Li$_{3}$BN$_{2}$ is 
0.81 eV \cite{yamane1987high} which is close to the above value of ${\Delta}E$(2N/4N), suggesting
that one possible mechanism of Li-ion conduction in Li$_{x}$BN$_{2}$ (1$\leq$x$\leq$3) is based on Li$^{+}$
jumping between nearby Li(2N) to Li(4N) sites, with the Li(4N) site being the transition state.
The present study investigated other possible mechanisms as well, such as Li(4N) to Li(4N) jumping parallel to
-Li-N-B-N- planes or perpendicular to these planes. In all cases the transition states required much
higher energies (2 eV and above) as they induced large geometric changes.

Partial (L\"owdin) charges of the various types of atoms are listed in Table \ref{bindingE}.
Note that the charges do not sum up to the formal charges of the corresponding ions (such as
BN$_{2}^{3-}$ and BN$_{2}^{2-}$) since the projection from plane-wave basis into atomic orbitals
is incomplete. However, tendencies of atomic charges give an account on how the charge is stored in
the various oxidation states of the system. As the oxidation of BN$_{2}^{n-}$ (1$\leq$n$\leq$3) goes on with
decreasing n and decreasing Li-content, the B atoms become gradually slightly 
more positively charged while the N atoms become significantly 
less negatively charged and the charge of the Li-s stays constant. 
The fact that B has near zero net charge in
BN$_{2}^{3-}$ and in its oxidized forms indicates that B carries significantly more electrons in these
ions than it does in h-BN, where the charge of B is $\approx$ + 0.5.
From this analysis
it is clear that the reduction/oxidation processes in the Li$_{x}$BN$_{2}$ (1$\leq$x$\leq$3) system are
mainly associated with the N atoms, in concert with the above mentioned analogies of Li$_{x}$BN$_{2}$ 
to various types of
molecules with varying oxidation number of the N atom (Li$_{3}$N, N$_{2}$H$_{4}$, Li$_{2}$N$_{2}$).

The application of the $\beta$ and monoclinic phases as electroactive material
is expected to result in large
volume changes, albeit at similarly high voltages, as in all cases the BN$_{2}^{3-}$ ions are
oxidized/reduced.
As the electrochemical cycling of batteries often causes phase transformations in the electroactive
crystals, it can be expected that the use of the $\beta$ and monoclinic phases could result in phase
transformation to the $\alpha$ phase whereby minimizing the volume changes of the electroactive
material. 

\section{Summary and Conclusions}
Despite the extensive research on hydrogen storage materials, such as Li$_{3}$BN$_{2}$H$_{8}$ 
\cite{pinkerton2006tetragonal} that decompose to Li$_{3}$BN$_{2}$ when heated, Li$_{3}$BN$_{2}$ has
not been tested as electroactive species in the positive electrodes of batteries, yet. It has been considered
though as component of a conversion based negative electrode material \cite{mason2011first}.
The present work provides a detailed theoretical analysis on the use of $\alpha$-Li$_{x}$BN$_{2}$
(1$\leq$x$\leq$3) as electroactive species in the positive electrode of electrochemical cells.
It is predicted by means of density functional theory calculations, that the application of
this material can result in 3.62 V cells (relative to Li/Li$^{+}$), a gravimetric energy density of
3251 Wh/kg, a volumetric one of 5927 Wh/L and gravimetric and volumetric capacities of
899 mAh/g and 1638 mAh/cm$^{3}$, respectively, when two Li ions are intercalated per formula unit.  
The predicted associated volume change is only 2.8\% per two-electron
transfer. The Li-deintercalated structures are metallic, when sufficient amount of Li is extracted.
These predicted properties are far superior to other existing or designer Li-intercalation based
battery materials and are even comparable to the theoretical energy density of a Li-O$_{2}$/peroxide
battery, 3450 Wh/kg, when O$_{2}$ is carried within the battery.
Furthermore, the $\alpha$-Li$_{y}$BN$_{2}$ (0$\leq$y$\leq$3) materials are also interesting as 
they are based on novel 1D conjugated $\pi$-electron systems representing new conducting polymers.

\section{Acknowledgement}
The author thanks Prof. L. Shaw (IIT) for discussions and 
NERSC (U.S. DOE DE-AC02-05CH11231) for the use of computational resources. 

\footnotetext{\textit{$^{b}$~Address, Physics Department, Illinois Institute of Technology,
Chicago, Illinois 60616, USA, nemeth@agni.phys.iit.edu}}

\footnotesize{

\begin{mcitethebibliography}{39}
\providecommand*{\natexlab}[1]{#1}
\providecommand*{\mciteSetBstSublistMode}[1]{}
\providecommand*{\mciteSetBstMaxWidthForm}[2]{}
\providecommand*{\mciteBstWouldAddEndPuncttrue}
  {\def\EndOfBibitem{\unskip.}}
\providecommand*{\mciteBstWouldAddEndPunctfalse}
  {\let\EndOfBibitem\relax}
\providecommand*{\mciteSetBstMidEndSepPunct}[3]{}
\providecommand*{\mciteSetBstSublistLabelBeginEnd}[3]{}
\providecommand*{\EndOfBibitem}{}
\mciteSetBstSublistMode{f}
\mciteSetBstMaxWidthForm{subitem}
{(\emph{\alph{mcitesubitemcount}})}
\mciteSetBstSublistLabelBeginEnd{\mcitemaxwidthsubitemform\space}
{\relax}{\relax}

\bibitem[Xu \emph{et~al.}(2012)Xu, Qian, Wang, and Meng]{BXu12}
B.~Xu, D.~Qian, Z.~Wang and Y.~S. Meng, \emph{Materials Science and
  Engineering: R: Reports}, 2012, \textbf{73}, 51--65\relax
\mciteBstWouldAddEndPuncttrue
\mciteSetBstMidEndSepPunct{\mcitedefaultmidpunct}
{\mcitedefaultendpunct}{\mcitedefaultseppunct}\relax
\EndOfBibitem
\bibitem[Melot and Tarascon(2013)]{BCMelot13}
B.~C. Melot and J.-M. Tarascon, \emph{Acc. Chem. Res.}, 2013, \textbf{46},
  1226--1238\relax
\mciteBstWouldAddEndPuncttrue
\mciteSetBstMidEndSepPunct{\mcitedefaultmidpunct}
{\mcitedefaultendpunct}{\mcitedefaultseppunct}\relax
\EndOfBibitem
\bibitem[Song \emph{et~al.}(2013)Song, Zhang, and Cairns]{MKSong13}
M.-K. Song, Y.~Zhang and E.~J. Cairns, \emph{Nano letters}, 2013, \textbf{13},
  5891--5899\relax
\mciteBstWouldAddEndPuncttrue
\mciteSetBstMidEndSepPunct{\mcitedefaultmidpunct}
{\mcitedefaultendpunct}{\mcitedefaultseppunct}\relax
\EndOfBibitem
\bibitem[Meduri \emph{et~al.}(2013)Meduri, Chen, Xiao, Martinez, Carlson,
  Zhang, and Deng]{PMeduri13}
P.~Meduri, H.~Chen, J.~Xiao, J.~J. Martinez, T.~Carlson, J.-G. Zhang and Z.~D.
  Deng, \emph{Journal of Materials Chemistry A}, 2013, \textbf{1},
  7866--7869\relax
\mciteBstWouldAddEndPuncttrue
\mciteSetBstMidEndSepPunct{\mcitedefaultmidpunct}
{\mcitedefaultendpunct}{\mcitedefaultseppunct}\relax
\EndOfBibitem
\bibitem[Li \emph{et~al.}(2012)Li, Meng, and Jin]{LLi12}
L.~Li, F.~Meng and S.~Jin, \emph{Nano letters}, 2012, \textbf{12},
  6030--6037\relax
\mciteBstWouldAddEndPuncttrue
\mciteSetBstMidEndSepPunct{\mcitedefaultmidpunct}
{\mcitedefaultendpunct}{\mcitedefaultseppunct}\relax
\EndOfBibitem
\bibitem[Hautier \emph{et~al.}(2011)Hautier, Jain, Chen, Moore, Ong, and
  Ceder]{GHautier11}
G.~Hautier, A.~Jain, H.~Chen, C.~Moore, S.~P. Ong and G.~Ceder, \emph{J. Mater.
  Chem.}, 2011, \textbf{21}, 17147–17153\relax
\mciteBstWouldAddEndPuncttrue
\mciteSetBstMidEndSepPunct{\mcitedefaultmidpunct}
{\mcitedefaultendpunct}{\mcitedefaultseppunct}\relax
\EndOfBibitem
\bibitem[Thackeray and {et al.}(2012)]{MMThackeray12}
M.~M. Thackeray and {et al.}, \emph{{Lithium-Oxygen (Air) Electrochemical Cells
  and Batteries, Patent, US8313721}}, 2012\relax
\mciteBstWouldAddEndPuncttrue
\mciteSetBstMidEndSepPunct{\mcitedefaultmidpunct}
{\mcitedefaultendpunct}{\mcitedefaultseppunct}\relax
\EndOfBibitem
\bibitem[Christensen \emph{et~al.}(2012)Christensen, Albertus, Sanchez-Carrera,
  Lohmann, Kozinsky, Liedtke, Ahmed, and Kojic]{JChristensen12}
J.~Christensen, P.~Albertus, R.~S. Sanchez-Carrera, T.~Lohmann, B.~Kozinsky,
  R.~Liedtke, J.~Ahmed and A.~Kojic, \emph{Journal of The Electrochemical
  Society}, 2012, \textbf{159}, R1--R30\relax
\mciteBstWouldAddEndPuncttrue
\mciteSetBstMidEndSepPunct{\mcitedefaultmidpunct}
{\mcitedefaultendpunct}{\mcitedefaultseppunct}\relax
\EndOfBibitem
\bibitem[N{\'e}meth and Srajer(2014)]{KNemeth14co2}
K.~N{\'e}meth and G.~Srajer, \emph{RSC Advances}, 2014, \textbf{4}, 1879\relax
\mciteBstWouldAddEndPuncttrue
\mciteSetBstMidEndSepPunct{\mcitedefaultmidpunct}
{\mcitedefaultendpunct}{\mcitedefaultseppunct}\relax
\EndOfBibitem
\bibitem[Nov{\'a}k \emph{et~al.}(1997)Nov{\'a}k, M{\"u}ller, Santhanam, and
  Haas]{novak1997electrochemically}
P.~Nov{\'a}k, K.~M{\"u}ller, K.~Santhanam and O.~Haas, \emph{Chemical Reviews},
  1997, \textbf{97}, 207--282\relax
\mciteBstWouldAddEndPuncttrue
\mciteSetBstMidEndSepPunct{\mcitedefaultmidpunct}
{\mcitedefaultendpunct}{\mcitedefaultseppunct}\relax
\EndOfBibitem
\bibitem[Archer(2004)]{archer2004inorganic}
R.~Archer, \emph{Inorganic and Organometallic Polymers}, Wiley, 2004\relax
\mciteBstWouldAddEndPuncttrue
\mciteSetBstMidEndSepPunct{\mcitedefaultmidpunct}
{\mcitedefaultendpunct}{\mcitedefaultseppunct}\relax
\EndOfBibitem
\bibitem[Downs \emph{et~al.}(1980)Downs, Fair, and Iqbal]{downs1980electric}
D.~Downs, H.~Fair and Z.~Iqbal, \emph{Electric initiator containing polymeric
  sulfur nitride}, 1980, {US Patent} 4,206,705\relax
\mciteBstWouldAddEndPuncttrue
\mciteSetBstMidEndSepPunct{\mcitedefaultmidpunct}
{\mcitedefaultendpunct}{\mcitedefaultseppunct}\relax
\EndOfBibitem
\bibitem[Ruschewitz(2006)]{ruschewitz2006ternary}
U.~Ruschewitz, \emph{Zeitschrift f{\"u}r anorganische und allgemeine Chemie},
  2006, \textbf{632}, 705--719\relax
\mciteBstWouldAddEndPuncttrue
\mciteSetBstMidEndSepPunct{\mcitedefaultmidpunct}
{\mcitedefaultendpunct}{\mcitedefaultseppunct}\relax
\EndOfBibitem
\bibitem[Terdik \emph{et~al.}(2012)Terdik, N{\'e}meth, Harkay, Terry~Jr,
  Spentzouris, Vel{\'a}zquez, Rosenberg, and Srajer]{terdik2012anomalous}
J.~Z. Terdik, K.~N{\'e}meth, K.~C. Harkay, J.~H. Terry~Jr, L.~Spentzouris,
  D.~Vel{\'a}zquez, R.~Rosenberg and G.~Srajer, \emph{Physical Review B}, 2012,
  \textbf{86}, 035142\relax
\mciteBstWouldAddEndPuncttrue
\mciteSetBstMidEndSepPunct{\mcitedefaultmidpunct}
{\mcitedefaultendpunct}{\mcitedefaultseppunct}\relax
\EndOfBibitem
\bibitem[Cenzual \emph{et~al.}(1991)Cenzual, Gelato, Penzo, and
  Parth{\'e}]{cenzual1991inorganic}
K.~Cenzual, L.~M. Gelato, M.~Penzo and E.~Parth{\'e}, \emph{Acta
  Crystallographica Section B: Structural Science}, 1991, \textbf{47},
  433--439\relax
\mciteBstWouldAddEndPuncttrue
\mciteSetBstMidEndSepPunct{\mcitedefaultmidpunct}
{\mcitedefaultendpunct}{\mcitedefaultseppunct}\relax
\EndOfBibitem
\bibitem[Yamane \emph{et~al.}(1987)Yamane, Kikkawa, and
  Koizumi]{yamane1987high}
H.~Yamane, S.~Kikkawa and M.~Koizumi, \emph{Journal of Solid State Chemistry},
  1987, \textbf{71}, 1--11\relax
\mciteBstWouldAddEndPuncttrue
\mciteSetBstMidEndSepPunct{\mcitedefaultmidpunct}
{\mcitedefaultendpunct}{\mcitedefaultseppunct}\relax
\EndOfBibitem
\bibitem[DeVries and Fleischer(1969)]{devries1969system}
R.~DeVries and J.~Fleischer, \emph{Materials Research Bulletin}, 1969,
  \textbf{4}, 433--441\relax
\mciteBstWouldAddEndPuncttrue
\mciteSetBstMidEndSepPunct{\mcitedefaultmidpunct}
{\mcitedefaultendpunct}{\mcitedefaultseppunct}\relax
\EndOfBibitem
\bibitem[Goubeau and Anselment(1961)]{goubeau1961ternare}
J.~Goubeau and W.~Anselment, \emph{Zeitschrift f{\"u}r anorganische und
  allgemeine Chemie}, 1961, \textbf{310}, 248--260\relax
\mciteBstWouldAddEndPuncttrue
\mciteSetBstMidEndSepPunct{\mcitedefaultmidpunct}
{\mcitedefaultendpunct}{\mcitedefaultseppunct}\relax
\EndOfBibitem
\bibitem[Pinkerton and Herbst(2006)]{pinkerton2006tetragonal}
F.~Pinkerton and J.~Herbst, \emph{Journal of applied physics}, 2006,
  \textbf{99}, 113523--113523\relax
\mciteBstWouldAddEndPuncttrue
\mciteSetBstMidEndSepPunct{\mcitedefaultmidpunct}
{\mcitedefaultendpunct}{\mcitedefaultseppunct}\relax
\EndOfBibitem
\bibitem[Yamane \emph{et~al.}(1986)Yamane, Kikkawa, Horiuchi, and
  Koizumi]{yamane1986structure}
H.~Yamane, S.~Kikkawa, H.~Horiuchi and M.~Koizumi, \emph{Journal of solid state
  chemistry}, 1986, \textbf{65}, 6--12\relax
\mciteBstWouldAddEndPuncttrue
\mciteSetBstMidEndSepPunct{\mcitedefaultmidpunct}
{\mcitedefaultendpunct}{\mcitedefaultseppunct}\relax
\EndOfBibitem
\bibitem[Yamane \emph{et~al.}(1987)Yamane, Kikkawa, and
  Koizumi]{yamane1987preparation}
H.~Yamane, S.~Kikkawa and M.~Koizumi, \emph{Journal of Power Sources}, 1987,
  \textbf{20}, 311--315\relax
\mciteBstWouldAddEndPuncttrue
\mciteSetBstMidEndSepPunct{\mcitedefaultmidpunct}
{\mcitedefaultendpunct}{\mcitedefaultseppunct}\relax
\EndOfBibitem
\bibitem[Qiu \emph{et~al.}(2012)Qiu, Zhuang, Zhang, Cao, Ying, Qiang, and
  Sun]{qiu2012electrochemical}
X.-Y. Qiu, Q.-C. Zhuang, Q.-Q. Zhang, R.~Cao, P.-Z. Ying, Y.-H. Qiang and S.-G.
  Sun, \emph{Physical Chemistry Chemical Physics}, 2012, \textbf{14},
  2617--2630\relax
\mciteBstWouldAddEndPuncttrue
\mciteSetBstMidEndSepPunct{\mcitedefaultmidpunct}
{\mcitedefaultendpunct}{\mcitedefaultseppunct}\relax
\EndOfBibitem
\bibitem[Waechter and Nesper(2013)]{waechter2013coating}
F.~Waechter and R.~Nesper, \emph{Coating and lithiation of inorganic oxidants
  by reaction with lithiated reductants}, 2013, {US Patent App.}
  13/724,748\relax
\mciteBstWouldAddEndPuncttrue
\mciteSetBstMidEndSepPunct{\mcitedefaultmidpunct}
{\mcitedefaultendpunct}{\mcitedefaultseppunct}\relax
\EndOfBibitem
\bibitem[Mason \emph{et~al.}(2011)Mason, Liu, Hong, Graetz, and
  Majzoub]{mason2011first}
T.~H. Mason, X.~Liu, J.~Hong, J.~Graetz and E.~Majzoub, \emph{The Journal of
  Physical Chemistry C}, 2011, \textbf{115}, 16681--16687\relax
\mciteBstWouldAddEndPuncttrue
\mciteSetBstMidEndSepPunct{\mcitedefaultmidpunct}
{\mcitedefaultendpunct}{\mcitedefaultseppunct}\relax
\EndOfBibitem
\bibitem[N\'emeth(2014)]{KNemeth2014ijqc}
K.~N\'emeth, \emph{International Journal of Quantum Chemistry}, 2014,
  \textbf{114}, in press\relax
\mciteBstWouldAddEndPuncttrue
\mciteSetBstMidEndSepPunct{\mcitedefaultmidpunct}
{\mcitedefaultendpunct}{\mcitedefaultseppunct}\relax
\EndOfBibitem
\bibitem[Zhou \emph{et~al.}(2004)Zhou, Cococcioni, Kang, and Ceder]{FZhou04}
F.~Zhou, M.~Cococcioni, K.~Kang and G.~Ceder, \emph{Electrochem. Comm.}, 2004,
  \textbf{6}, 1144--1148\relax
\mciteBstWouldAddEndPuncttrue
\mciteSetBstMidEndSepPunct{\mcitedefaultmidpunct}
{\mcitedefaultendpunct}{\mcitedefaultseppunct}\relax
\EndOfBibitem
\bibitem[Jain \emph{et~al.}(2011)Jain, Hautier, Ong, Moore, Fischer, Persson,
  and Ceder]{AJain11}
A.~Jain, G.~Hautier, S.~P. Ong, C.~J. Moore, C.~C. Fischer, K.~A. Persson and
  G.~Ceder, \emph{Phys. Rev. B}, 2011, \textbf{84}, 045115\relax
\mciteBstWouldAddEndPuncttrue
\mciteSetBstMidEndSepPunct{\mcitedefaultmidpunct}
{\mcitedefaultendpunct}{\mcitedefaultseppunct}\relax
\EndOfBibitem
\bibitem[Anisimov \emph{et~al.}(1991)Anisimov, Zaanen, and
  Andersen]{anisimov1991band}
V.~I. Anisimov, J.~Zaanen and O.~K. Andersen, \emph{Physical Review B}, 1991,
  \textbf{44}, 943\relax
\mciteBstWouldAddEndPuncttrue
\mciteSetBstMidEndSepPunct{\mcitedefaultmidpunct}
{\mcitedefaultendpunct}{\mcitedefaultseppunct}\relax
\EndOfBibitem
\bibitem[Perdew \emph{et~al.}(1996)Perdew, Burke, and Ernzerhof]{PBE}
J.~P. Perdew, K.~Burke and M.~Ernzerhof, \emph{Phys. Rev. Lett.}, 1996,
  \textbf{77}, 3865\relax
\mciteBstWouldAddEndPuncttrue
\mciteSetBstMidEndSepPunct{\mcitedefaultmidpunct}
{\mcitedefaultendpunct}{\mcitedefaultseppunct}\relax
\EndOfBibitem
\bibitem[Perdew \emph{et~al.}(2008)Perdew, Ruzsinszky, Csonka, Vydrov,
  Scuseria, Constantin, Zhou, and Burke]{PBEsol}
J.~P. Perdew, A.~Ruzsinszky, G.~I. Csonka, O.~A. Vydrov, G.~E. Scuseria, L.~A.
  Constantin, X.~Zhou and K.~Burke, \emph{Phys. Rev. Lett.}, 2008,
  \textbf{100}, 136406\relax
\mciteBstWouldAddEndPuncttrue
\mciteSetBstMidEndSepPunct{\mcitedefaultmidpunct}
{\mcitedefaultendpunct}{\mcitedefaultseppunct}\relax
\EndOfBibitem
\bibitem[Giannozzi \emph{et~al.}(2009)Giannozzi, Baroni, Bonini, Calandra, Car,
  Cavazzoni, Ceresoli, Chiarotti, Cococcioni,
  Dabo,\emph{et~al.}]{giannozzi2009quantum}
P.~Giannozzi, S.~Baroni, N.~Bonini, M.~Calandra, R.~Car, C.~Cavazzoni,
  D.~Ceresoli, G.~L. Chiarotti, M.~Cococcioni, I.~Dabo \emph{et~al.},
  \emph{Journal of Physics: Condensed Matter}, 2009, \textbf{21}, 395502\relax
\mciteBstWouldAddEndPuncttrue
\mciteSetBstMidEndSepPunct{\mcitedefaultmidpunct}
{\mcitedefaultendpunct}{\mcitedefaultseppunct}\relax
\EndOfBibitem
\bibitem[Huq \emph{et~al.}(2007)Huq, Richardson, Maxey, Chandra, and
  Chien]{huq2007structural}
A.~Huq, J.~W. Richardson, E.~R. Maxey, D.~Chandra and W.-M. Chien,
  \emph{Journal of alloys and compounds}, 2007, \textbf{436}, 256--260\relax
\mciteBstWouldAddEndPuncttrue
\mciteSetBstMidEndSepPunct{\mcitedefaultmidpunct}
{\mcitedefaultendpunct}{\mcitedefaultseppunct}\relax
\EndOfBibitem
\bibitem[Pease(1952)]{pease1952x}
R.~S. Pease, \emph{Acta Crystallographica}, 1952, \textbf{5}, 356--361\relax
\mciteBstWouldAddEndPuncttrue
\mciteSetBstMidEndSepPunct{\mcitedefaultmidpunct}
{\mcitedefaultendpunct}{\mcitedefaultseppunct}\relax
\EndOfBibitem
\bibitem[McHale \emph{et~al.}(1999)McHale, Navrotsky, and
  DiSalvo]{mchale1999energetics}
J.~McHale, A.~Navrotsky and F.~DiSalvo, \emph{Chemistry of materials}, 1999,
  \textbf{11}, 1148--1152\relax
\mciteBstWouldAddEndPuncttrue
\mciteSetBstMidEndSepPunct{\mcitedefaultmidpunct}
{\mcitedefaultendpunct}{\mcitedefaultseppunct}\relax
\EndOfBibitem
\bibitem[{Jr. Chase}(1998)]{MWJrChase98}
\emph{NIST-JANAF Themochemical Tables}, ed. M.~W. {Jr. Chase}, American
  Chemical Society and the American Institute of Physics, Washington DC, 4th
  edn, 1998, pp. 1--1951\relax
\mciteBstWouldAddEndPuncttrue
\mciteSetBstMidEndSepPunct{\mcitedefaultmidpunct}
{\mcitedefaultendpunct}{\mcitedefaultseppunct}\relax
\EndOfBibitem
\bibitem[Wise \emph{et~al.}(1966)Wise, Margrave, Feder, and
  Hubbard]{wise1966fluorine}
S.~S. Wise, J.~L. Margrave, H.~M. Feder and W.~N. Hubbard, \emph{The Journal of
  Physical Chemistry}, 1966, \textbf{70}, 7--10\relax
\mciteBstWouldAddEndPuncttrue
\mciteSetBstMidEndSepPunct{\mcitedefaultmidpunct}
{\mcitedefaultendpunct}{\mcitedefaultseppunct}\relax
\EndOfBibitem
\bibitem[Blaschkowski(2003)]{Blaschkowski03}
B.~Blaschkowski, \emph{PhD thesis}, Universit\"at T\"ubingen, Wilhelmstr. 32,
  72074 T\"ubingen, 2003\relax
\mciteBstWouldAddEndPuncttrue
\mciteSetBstMidEndSepPunct{\mcitedefaultmidpunct}
{\mcitedefaultendpunct}{\mcitedefaultseppunct}\relax
\EndOfBibitem
\bibitem[Schneider \emph{et~al.}(2012)Schneider, Frankovsky, and
  Schnick]{schneider2012high}
S.~B. Schneider, R.~Frankovsky and W.~Schnick, \emph{Angewandte Chemie}, 2012,
  \textbf{124}, 1909--1911\relax
\mciteBstWouldAddEndPuncttrue
\mciteSetBstMidEndSepPunct{\mcitedefaultmidpunct}
{\mcitedefaultendpunct}{\mcitedefaultseppunct}\relax
\EndOfBibitem
\bibitem[Koz \emph{et~al.}(2014)Koz, Acar, Prots, H{\"o}hn, and
  Somer]{koz2014na3}
C.~Koz, S.~Acar, Y.~Prots, P.~H{\"o}hn and M.~Somer, \emph{Zeitschrift f{\"u}r
  anorganische und allgemeine Chemie}, 2014\relax
\mciteBstWouldAddEndPuncttrue
\mciteSetBstMidEndSepPunct{\mcitedefaultmidpunct}
{\mcitedefaultendpunct}{\mcitedefaultseppunct}\relax
\EndOfBibitem
\end{mcitethebibliography}
\bibliographystyle{rsc} 
\providecommand*{\mcitethebibliography}{\thebibliography}
\csname @ifundefined\endcsname{endmcitethebibliography}
{\let\endmcitethebibliography\endthebibliography}{}

}
\noindent

\end{document}